\documentclass[twocolumn,showpacs,preprintnumbers,nofootinbib,amsmath,amssymb,floatfix,aps]{revtex4-1}
\usepackage{dcolumn}
\usepackage{bm}
\usepackage{color}

\usepackage{graphicx}
\usepackage{amsmath}
\usepackage{amssymb}

\def\beq{\begin{equation}}
\def\eeq{\end{equation}}
\def\be{\begin{eqnarray}}
\def\ee{\end{eqnarray}}

\begin{document}
\title{Interpretation of Neutrino-Nucleus Cross Sections}
\author{Omar Benhar}
\affiliation{INFN and Department of Physics, ``Sapienza'' University, I-00185 Roma, Italy,}
\affiliation{Center for Neutrino Physics, Virginia Tech, Blacksburg, Virginia 24061, USA}

\date{\today}
\begin{abstract}
I discuss the near-degeneracy between models of neutrino-nucleus interactions based on diverse assumptions, and analyze a specific example illustrating how the different reaction 
mechanisms taken into account, as well as the approximations associated with their implementation,  
may conspire to make the prediction of two models look similar. I argue that the relevant reaction mechanisms may be unambiguously identified
exploiting the availability of electron-nucleus scattering data other than inclusive cross sections, providing detailed information on the nuclear spectral function.

\end{abstract}
\maketitle

\section{Introduction}

{\em Varius, multiplex, multiformis}.  The three latin adjectives famously employed to describe the complex personality of roman emperor Hadrian \cite{memoirs}  
provide a remarkably accurate characterization of the nuclear response to neutrino interactions, resulting from the combination of 
{\em many} and 
{\em diverse} mechanisms, giving rise to {\em distinct} observable effects that hamper the interpretation of the measured cross sections.

The interest in theoretical modeling of neutrino-nucleus scattering was suddenly revived in the early 2000s, when 
the treatment of nuclear effects was clearly recognized as a major source of systematic error for accelerator-based 
searches 
of neutrino oscillations~\cite{NUINT}. Few years later, the inadequacy of the Relativistic Fermi Gas Model (RFMG), largely employed in Monte Carlo simulations,   
was clearly exposed by its conspicuous failure to explain the double-differential  $\nu_\mu$-carbon cross section
in the charged-current quasi elastic channel, measured by the MiniBooNE Collaboration \cite{CCQE}.

The studies carried out over the past decade have led to the development of a number of more advanced 
models of quasi elastic neutrino-nucleus scattering, taking into account 
both the effects of strong interaction dynamics and the variety of mechanisms contributing to the flux integrated cross section~\cite{benhar_PRD,coletti,ankowski,rocco,martini,nieves,SuSa,GIBBU,natalie}, 
some of which have reached the
degree of maturity required for a meaningful comparison between their predictions and the available data.
In this context, an essential role has been played by the availability of a large body of electron-nucleus
scattering cross sections, precisely measured using a variety of targets and spanning a broad kinematical range. While the description of the measured neutrino-nucleus cross sections  
involves non trivial additional problems, mainly arising from the flux average~\cite{benhar_nufact11}, the ability to explain electron scattering data must in fact be regarded as an obvious prerequisite, to be met by any models of the nuclear response to weak interactions. 

Considerable progress has been also achieved by the {\em ab initio} Quantum Monte Carlo method,  which allows to carry out very accurate calculations of the electromagnetic and weak responses of 
nuclei as heavy as carbon~\cite{GFMC1}. This approach provides a
remarkably good account of inclusive electron-nucleus scattering data in the kinematical regime in which the non relativistic approximation is expected to be applicable~\cite{GFMC2}. 

In many instances, the predictions of theoretical models turn out to be in satisfactory agreement with experiments. However,  a deeper
scrutiny reveals a puzzling feature: models based on conceptually different\textemdash sometimes even contradictory\textemdash
assumptions yield comparable results. In view of applications to the different kinematical regimes and nuclear targets relevant to 
future experiments, the sources of degeneracy between different theoretical approaches\textemdash implying that the agreement between theory and data may 
in fact be accidental\textemdash need to be identified, so as to firmly assess their predictive power.

Section \ref{sec2}  is devoted to the analysis of a specific example, illustrating how the different reaction mechanisms 
included in different theoretical models, as well as the approximations associated with their implementation,  
may conspire to make the predicted cross sections look similar. In Section \ref{sec3}, I discuss the uncertainties associated with the identification of single-nucleon emission 
processes\textemdash yielding the largest contribution to the quasi elastic cross section at momentum transfer in the hundred-MeV range\textemdash and argue that they 
may be substantially reduced
exploiting available electron scattering data. Finally, in Section \ref{sec4} I state the conclusions, and outline a possible 
avenue for the development of a comprehensive and consistent description of neutrino-nucleus scattering. 

\section{The degeneracy issue}
\label{sec2}

The differential cross section of the process 
\begin{align}
\nu_\mu + A \to \mu^- + X \   , 
\end{align}
where $A$ and $X$ denote the target nucleus
in its ground state and the hadronic final state, respectively, 
can be schematically written in the form 
\begin{align}
\nonumber
{d \sigma}_A \propto L_{\alpha \beta} \sum_X  & \left[ \langle0 |  {J^\alpha_A}^\dagger  | X \rangle 
                                                            \langle X |  {J^\beta_A} | 0 \rangle + h.c. \right] \\
                                                            & \times \delta^{(4)}(P_0 + q - P_X) \ , 
\label{nucl:xsec}
\end{align}
with the tensor $L_{\alpha \beta}$ being  fully specified by the lepton kinematical variables. The above equation shows that
the description of the nuclear response involves the target initial and final states, carrying four-momenta $P_0$ and $P_X$, as well as the nuclear current operator
\begin{align}
 {J^\alpha_A} = \sum_i {j^\alpha_i} + \sum_{j>i} {j^\alpha_{ij}} \ , 
\end{align} 
comprising one- and two-nucleon terms. 
The sum in Eq.~\eqref{nucl:xsec} includes contributions from all possible final states, excited through different reaction mechanisms whose relative weight  
depends on kinematics.

Charged Current Quasi elastic (CCQE) scattering off an individual nucleon, that is the process corresponding to the final state
\begin{align}
\label{1p1h}
| X \rangle = | p, (A-1)^*  \rangle \ , 
\end{align}
is the dominant mechanism in the kinematical region relevant to the analysis of the MiniBooNE data, corresponding 
to a neutrino flux with mean energy $\langle E_\nu \rangle = 880 \ {\rm MeV}$. From the experimental point of view, CCQE events 
are characterized by the absence of pions in the final state. They are identified  
by the measured kinetic energy and emission angle of the muon, with the knocked out proton and the recoiling nucleus being undetected. 
Note that the spectator $(A-1)$-nucleon system can either be in a bound state
or include a nucleon excited to the continuum\footnote{Theoretical studies of the momentum distribution sum rule in isospin-symmetric nuclear matter strongly suggest that 
the contribution of $(A-1)$-nucleon states involving more than one particle in the continuum is negligibly small~\cite{BFF}.}. 
For example, in the case of a carbon target the state of the recoiling
system can be $| ^{11}{\rm C}  \rangle$,  $| p , {^{10}{\rm B}}   \rangle$ or $|  n,  {^{10}{\rm C} }  \rangle$, and the corresponding $A$-nucleon final states are
\begin{align}
\label{1p1h_C}
| X \rangle = | p,  {^{11}{\rm C}}  \rangle  \ ,
\end{align}
or
\begin{align}
\label{2p2h_C}
| X \rangle = | p p,  {^{10}{\rm B}}  \rangle \ \ , \ \  | p n,  {^{10}{\rm C}}  \rangle  \ .
\end{align}
The states in the righ-hand side of Eqs.~\eqref{1p1h_C} and  \eqref{2p2h_C} are referred to as 
one-particle\textendash one-hole (1p1h) and two-particle\textendash two-hole (2p2h) states, respectively.

The appearance of 2p2h final states in scattering processes in which the beam particle couples to an individual nucleon originates   
from nucleon-nucleon correlations in the target ground state and final state interactions (FSI) between the struck particle and the spectator nucleons. 
These mechanisms are not taken into account by models based on the independent particle picture of the nucleus, such as the RFGM, according to 
which single nucleon knock out  can only lead to transitions to 1p1h final states. However, transitions to 2p2h states are always allowed in processes driven by two-nucleon 
meson-exchange currents (MEC), such as those 
in which the beam particle couples to a $\pi$-meson exchanged between two interacting nucleons.
A detailed discussion of the contributions of 1p1h and 2p2h final states to the nuclear response can be found in Ref.~\cite{BLR}.

More complex final states, that can be written as a superposition of 1p1h states according to 
\begin{align}
\label{RPA_C}
| X \rangle = \sum_n C_n | p_n h_n \rangle  \ ,
\end{align}
appear in processes in which the momentum transfer is shared between many nucleons.
The contribution of these processes is often described within the Random Phase Approximation  (RPA), 
which amounts to taking into account the so-called ring diagrams to all orders using phenomenological effective interactions~ \cite{RPA1,RPA2}.
On the basis of very general quantum-mechanical considerations,  long-range correlations associated with the final states of Eq.~\eqref{RPA_C} are expected to become 
important in the kinematical region in which the space resolution of the beam particle is much larger than the average 
nucleon-nucleon distance in the nuclear target, $d$, i.e. for typical momentum transfers $|{\bf q}| \ll  \pi/d \sim 400  \ {\rm MeV}$.

\begin{figure}[h!]
\vspace*{-.05in}
\begin{center}
\includegraphics[scale=0.7]{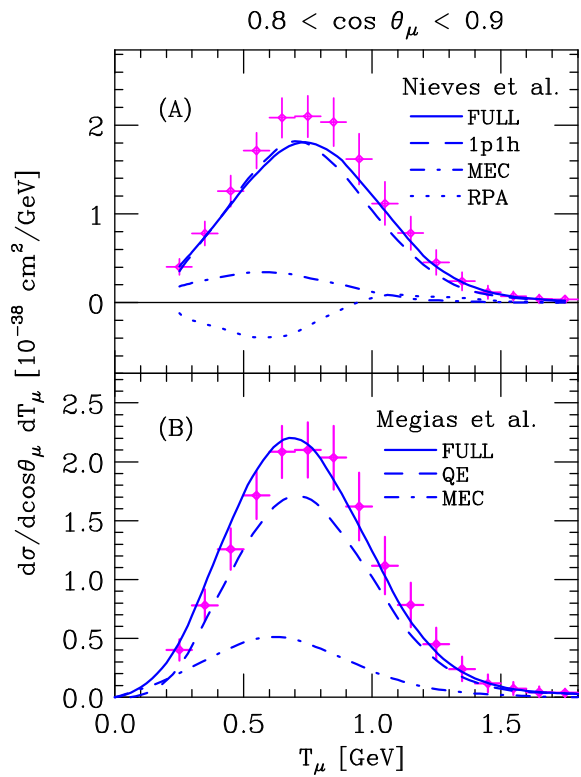}
\end{center}
\vspace*{-.20in}
\caption{Comparison between the 
double differential $\nu_\mu$-carbon cross section in the CCQE channel measured by the 
MiniBooNE Collaboration~\cite{CCQE} and the 
results obtained from the models of Nieves {\em et al.}~\cite{nieves} (A), and Megias {\em et al.}~\cite{SuSa}  (B). The solid lines 
correspond to the full calculations. The meaning of the dashed, dot-dash and dotted lines is explained 
in the text.}    
\label{degeneracy}
\end{figure}

The role played by the reaction mechanisms taken into account by two different models of neutino-nucleus interactions is illustrated in Fig.~\ref{degeneracy}, 
showing a comparison between the double-differential CCQE cross section measured by the MiniBooNE Collaboration~\cite{CCQE}
and the theoretical results reported in Refs.~\cite{nieves} [panel (A)] and \cite{SuSa} [panel (B)].

Within the approach of Ref.\cite{nieves}, the contribution of transitions to 1p1h final states\textemdash  described using the 
local Fermi gas approximation\textemdash is supplemented with those arising from processes involving MEC and 
long-range RPA correlations, obtained from the model originally proposed in Ref.~\cite{RPA2}. 

The hybrid model of Ref.~\cite{SuSa} combines the results of the phenomenological scaling analysis of electron scattering 
data \cite{yscaling}\textemdash extended to include inelastic channels\textemdash with a theoretical calculation of MEC contributions carried out 
within the RFGM. 

Overall, Fig.~\ref{degeneracy} shows that, 
up to a 10\% normalization uncertainty \cite{nieves}, the two models provide comparable
descriptions of the data. However, their predictions result from the combination of different reaction mechanisms.

Panel (A) indicates that, according to the model of Ref.~\cite{nieves}, the calculation including 1p1h final states only (dashed line labelled 1p1h) yields
a good approximation to the full result, represented by the solid line. The corrections arising from MEC (dot-dash line)
and long range correlations (dotted line labelled RPA) turn out to largely cancel one another. On the other hand, panel (B) suggests that, once 
single-nucleon knock out processes (dashed line labelled QE) and MEC (dot-dash line) are taken into account, the addition of long-range correlations\textemdash whose contribution {\em does not} exhibit scaling\textemdash  
is not needed to explain the data.  

\section{Single-nucleon emission}
\label{sec3}

The picture emerging from Fig.~\ref{degeneracy} clearly calls for a deeper analysis. As a first step, it is very important to realize that 
studying the role of mechanisms more complex than 
the excitation of 1p1h final states is only relevant to the extent to which the 1p1h sector, providing the dominant contribution to the cross section, is fully under control.
In this context, the results shown in Fig.~\ref{degeneracy} are not very useful. 

The 1p1h contribution of Ref.~\cite{nieves} [dashed line of panel (A)] is obtained within the independent particle picture of 
the nucleus, according to which all single-nucleon states belonging to the Fermi sea are occupied with unit probability.
However, the data collected over fifty years of $(e,e^\prime p)$ experiments~\cite{mougey:review,benhar:npn} have unambiguously demonstrated that the occupation probability of shell model states
is in fact sizably reduced\textemdash by as much as $\sim30-35 \%$ in the case of valence states\textemdash  by correlation effects.

The dashed line labelled QE in panel (B) also fails to provide an accurate estimate of the cross section in the 1p1h sector, because the empirical scaling function 
 includes additional contributions from processes involving the excitation of 2p2h final states, driven by ground state correlations, which are known to be non negligible~\cite{BLR}.

Detailed information on single nucleon knock out processes leading to the excitation of 1p1h final states has been obtained
studying the reactions
\begin{align}
e + A \to e^\prime + p + (A-1)_{\rm B} \ , 
\label{eep}
\end{align}
in which the scattered electron and the outgoing proton 
are detected in coincidence, and the recoiling nucleus is left in a bound state.
In the absence of final state interactions (FSI), whose effects can be taken into account as corrections, the $(e,e^\prime p)$ cross 
section reduces to the simple and transparent form
\begin{align}
d \sigma_{\rm eA} = \frac{|{\bf p}|}{ T_{p} + m} P(p_m, E_m) \ d \sigma_{ep} \ ,
\label{eep:xsec}
\end{align}
with the missing momentum and missing energy defined in terms of {\em measured} kinematical quantities as
\begin{align}
\label{eep:kin}
p_m = | {\bf p} - {\bf q} |  \ \ \ , \ \ \ E_m = \omega - T_{p} - T_{A-1}   \ . 
\end{align}
In the above equations, $\omega$ is the energy transfer, ${\bf p}$ and $T_{p}$ denote the momentum and kinetic energy of the emitted proton, respectively, 
and  $T_{A-1} = p_m^2/2M_{A-1}$  is the kinetic energy of the residual nucleus of mass $M_{A-1}$.

Equation~\eqref{eep:xsec} shows that a measurement of the $(e,e^\prime p)$  cross section 
gives access to the spectral function $P(p_m, E_m)$, describing the probability of removing a nucleon of momentum 
$p_m$ from the target nucleus leaving the residual system with excitation energy $E_m$. 

Being trivially related to the  imaginary part of the two-point Green's function, the spectral function admits an {\em exact} decomposition
into pole and continuum contributions~\cite{BFFZ}\textemdash known as K\"all\'en-Lehman representation\textemdash allowing a model independent 
identification of single nucleon emission processes, such as those of Eq.~\eqref{eep}, associated with 1p1h final states. From the experimental point of view, 
these reactions are signaled by the presence of sharp spectroscopic lines in the missing energy spectra measured at low to moderate 
$p_m$ and $E_m$, typically $p_m \textless  300 \ {\rm MeV}$ and $E_m \textless 50 \ {\rm MeV}$.
  
\begin{figure}[h!]
\begin{center}
\includegraphics[scale=0.7]{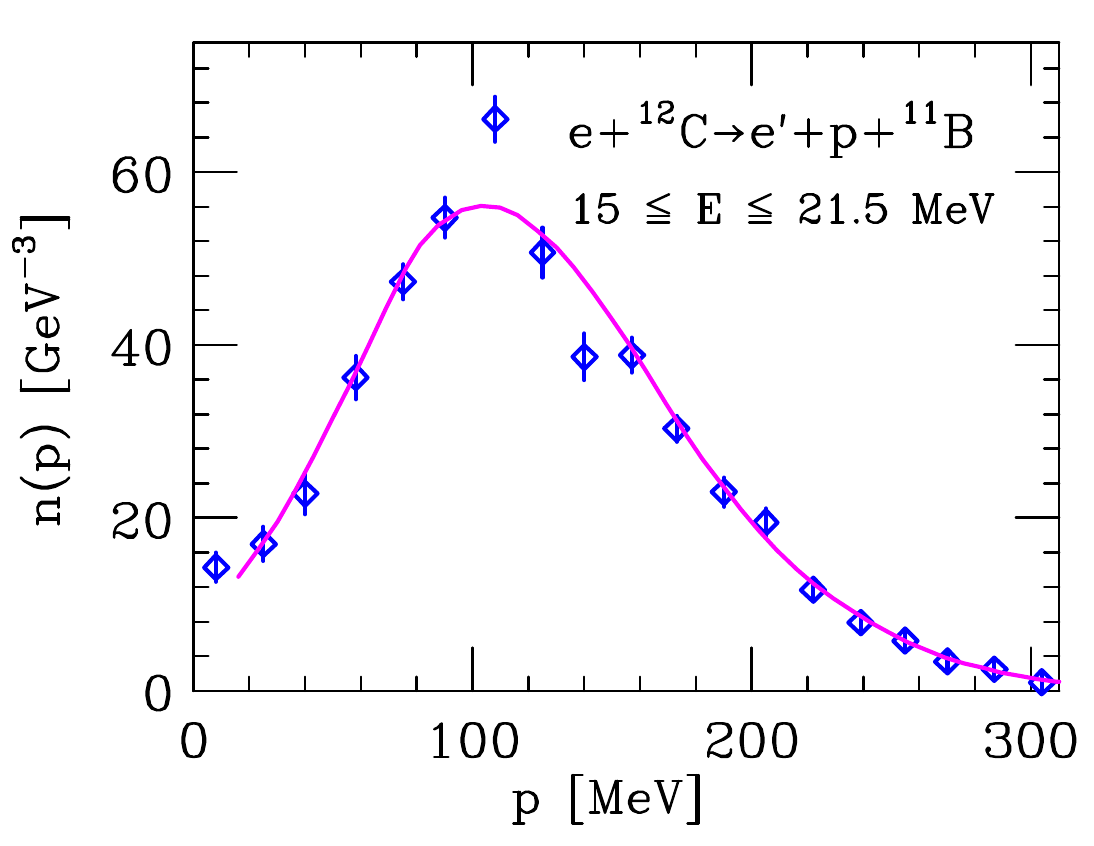}
\end{center}
\vspace*{-.2in}
\caption{Momentum distribution of the valence $P$-states of carbon, obtained from the $(e,e^\prime p)$ cross section 
of Ref.~\cite{mougey:76}. The solid line shows the momentum distribution obtained from the spectral function of
Ref.~\cite{LDA}, corrected to take into account the effects of final state interactions (FSI).} 
\label{momdis}
\end{figure}

Proton knock out from carbon in the kinematical region corresponding to single-nucleon emission has been  thoroughly investigated by 
Mougey {\em et al.} using the 
electron beam delivered by the Accelerateur Lineaire de Saclay (ALS) \cite{mougey:76}. The momentum distributions of the shell model states with quantum numbers specified by the index $\alpha$ ($\alpha = S, P$), defined as 
 \begin{align}
\label{momdis:def}
n_\alpha(p_m) =  \int_{E^\alpha_{\rm min}}^{E^\alpha_{\rm max}} P(p_m,E_m)  \ dE_m \ , 
\end{align} 
have been obtained using the spectral function extracted from the 
measured ${^{12}{\rm C}}(e,e^\prime p){^{11}{\rm B}}$ cross section, the integration regions being chosen
in such a way as to include the corresponding spectroscopic lines. 
As an example, Fig.~\ref{momdis} shows the momentum distribution of the valence P-states,   
computed  using Eq.~\eqref{momdis:def} with  $E^P_{\rm min}~= ~15 \ {\rm MeV}$ and $E^P_{\rm max}~=~22.5 \ {\rm MeV}$
~\cite{mougey:76}. Integration over $p_m$ yields the spectroscopic factor,  providing a measure of the occupation 
probability. The resulting value, $Z_P = 0.625$, implies that dynamical effects not taken into account within the 
independent particle model reduce the average number of P-state protons from 4 to 2.5.

The data of Ref.~\cite{mougey:76}
have been combined with the results of accurate theoretical calculation of the continuum component of the spectral function of isospin symmetric nuclear matter~\cite{BFF} 
to obtain the full carbon spectral function within the Local Density Approximation (LDA)~\cite{LDA}.

The P-state momentum  distribution computed from Eq.~\eqref{momdis:def} using the spectral function of Ref.~\cite{LDA}, corrected for FSI following the procedure of Ref.~\cite{mougey:76}, is shown by the solid line of Fig.~\ref{momdis}. It clearly appears that both shape and normalization are accurately accounted for. The spectroscopic
factor, $Z_P = 0.64$, turns out to be within $\sim2\%$ of the experimental value. 

A comparison with the data of Refs.~\cite{rohe1,transparency,rohe2}, reporting the results of a measurement of the $(e,e^\prime p)$ cross section at large $p_m$ and $E_m$ performed at the Thomas Jeffetson National 
Accelerator Facility (JLab), shows that 
the spectral function of Ref.~\cite{LDA} also provides a quantitative description of the contribution arising from nucleon-nucleon correlations. The continuum strength integrated over 
the region covered by the JLab experiment turns out to be $0.61\pm0.06$, to be compared with a theoretical value of 0.64~\cite{rohe2}.

\begin{figure}[h!]
\vspace*{.15in}
\begin{center}
\includegraphics[scale=0.7]{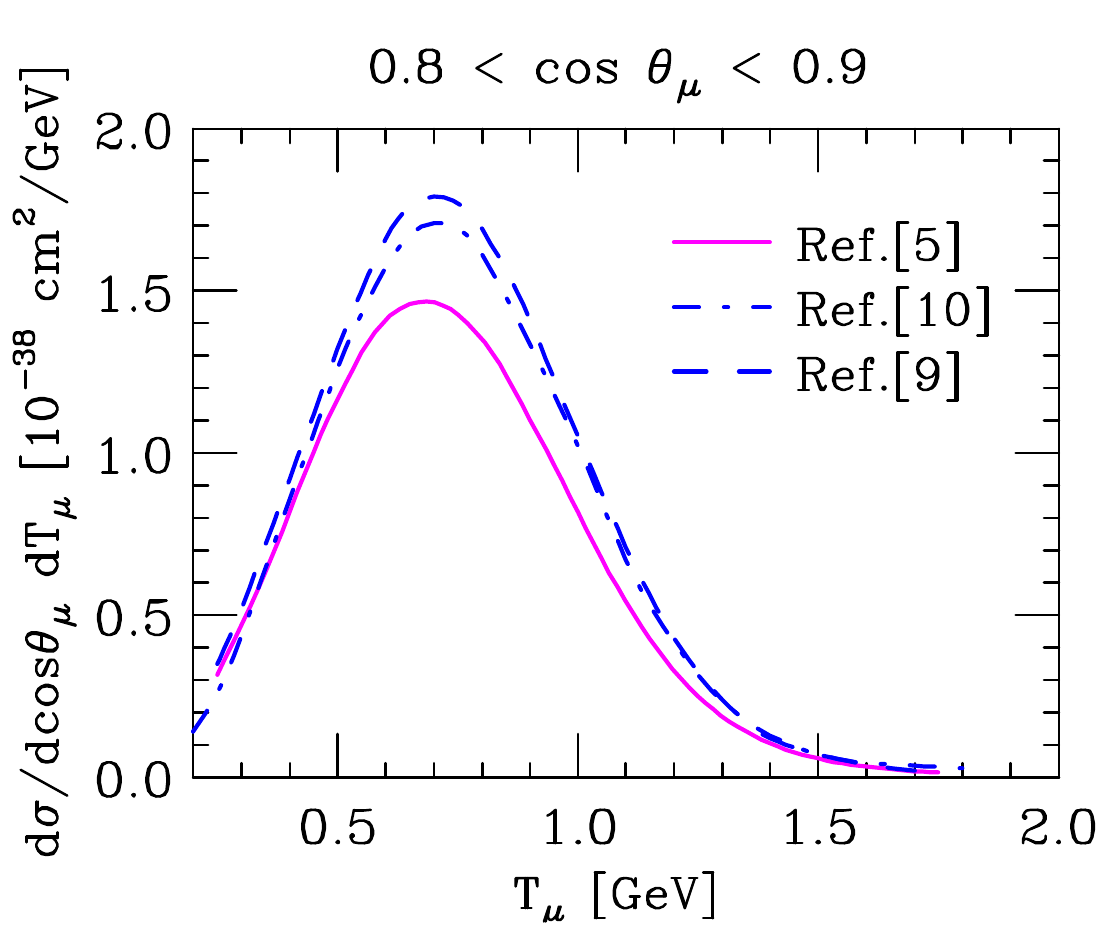}
\end{center}
\vspace*{-.30in}
\caption{Double differential $\nu_\mu$-carbon cross section in the CCQE channel, averaged over the MiniBooNe neutrino flux. The dashed and solid lines 
show the single nucleon emission contributions\textemdash corresponding to 1p1h final states\textemdash 
obtained from the model of Nieves {\em et al.}~\cite{nieves} and the spectral function formalism of Ref.~\cite{benhar_PRD,coletti}, 
respectively. The dot-dash line represents the quasi elastic cross section obtained from the scaling analysis of Ref.~\cite{SuSa}. }    
\label{degeneracy2}
\end{figure}

The pole component of the spectral function of Ref.~\cite{LDA} can be employed to obtain the 1p1h contribution to the flux integrated CCQE $\nu_\mu$-carbon cross section within the approach 
described in Re.~\cite{benhar_PRD,coletti}. In Fig.~\ref{degeneracy2} the results of this calculation are compared with the 1p1h cross section of Ref.~\cite{nieves} and the QE result of Ref.~\cite{SuSa}.
As was to be expected on the basis of the discussion of Section~\ref{sec2}, both the local Fermi gas model and the phenomenological scaling analysis significantly overpredict  the 1p1h cross section 
obtained using the spectral function of Ref.~\cite{LDA}, strongly constrained by $(e,e^\prime p)$ data. Note that the $\sim20\%$ difference at the peak is about the same size as the discrepancy between the MiniBooNE data 
and the results of Monte Carlo simulations based on the RFGM, that stirred a great deal of debate on the need to use an {\em effective axial mass} in modeling neutrino-nucleus interactions.

\section{summary and outlook}
\label{sec4}

In spite of the remarkable progress achieved over of the past decade, the available theoretical models of neutrino-nucleus interactions, while yielding 
a good description of several measured cross sections, fail to provide a fully unambiguous interpretation of the data.

In order to establish if and to what extent the agreement between theory and experiments is in fact accidental, the near-degeneracy between different approaches
must be resolved, by putting their predictions to test 
against both electron scattering data and the results of accurate theoretical calculations. 

The measured cross sections of {\em exclusive} $(e,e^\prime p)$  processes, whose analysis allows to isolate the contribution 
of single nucleon knock out processes leading to the excitation of 1p1h final states,  provide the ideal tool for gauging the ability 
of theoretical models to describe the dominant reaction mechanism in the kinematical region of MiniBooNE data.
In order to acquire additional information, needed for the interpretation of signals detected using the liquid argon technology,  the 
available dataset will soon be augmented with the ${^{40}}{\rm Ar}(e,e^\prime p){^{39}}{\rm Cl}$ cross section, to be measured 
at Jlab in 2017~\cite{JLab}.

On the theory side, an important role will be played by the Green's Function Monte Carlo (GFMC) approach. 
The electromagnetic and weak responses obtained from GFMC calculations\textemdash based on realistic nuclear Hamiltonians
and consistently derived current operators\textemdash are exact up to statistical errors. On the other hand, the GFMC technique 
is inherently non relativistic, and does not allow to single out the contributions of different final states. As a consequence, its applications are somewhat limited, and 
do not yield detailed information on the reaction mechanisms. 
Nevertheless, besides being highly valuable in their own right\textemdash e.g. for studies of low-energy supernova neutrinos~\cite{supernova}\textemdash GFMC results 
can be effectively used as benchmarks, to assess the accuracy of more approximate calculations in the 
non relativistic limit. It must be kept in mind, however, that GFMC results can be meaningfully compared only to results obtained  from approaches based on 
the {\em same} model of nuclear dynamics~\cite{RLB}. 

The analysis of Section~\ref{sec3} shows that the scheme based on the spectral function formalism and a realistic model of nuclear dynamics, 
which is known to provide a quantitative account of the body of inclusive electron-carbon cross sections in the quasi elastic sector~\cite{ankowski}, also fulfills the requirement of 
describing the cross section in exclusive channels. 

Recent calculations carried out extending the formalism to include the contributions 
of two-nucleon currents and inelastic interactions, discussed in Ref.~\cite{rocco}, indicate that that the spectral function approach\textemdash 
founded on sound theoretical arguments derived within the conceptual framework of the Impulse Approximation (IA)\textemdash has the potential to provide 
a {\em comprehensive} and {\em consistent} interpretation of the electron-nucleus cross section at momentum transfer above few hundreds MeV.
At the same time, these results suggest that in this kinematical regime, typical of the CCQE events collected by the MiniBooNE collaboration, reaction mechanisms not included in the 
IA picture, most notably long range correlation described within RPA, are unlikely to play an important role. This observation turns out to be consistent with the pattern emerging 
from Fig.~\ref{degeneracy}.

The generalization of the work of Ref~\cite{rocco} to the case of neutrino scattering in all relevant channels does not involve conceptual difficulties. 
The treatment of quasi elastic scattering and resonance production is discussed in Refs.~\cite{benhar_PRD,coletti,BM}, while the calculation of the Deep Inelastic Scattering (DIS) 
and two-nucleon current contributions is under way. 

As a final remark, it is worth mentioning that, being largely based un probability distributions, the spectral function formalism is well suited for implementation in 
Monte Carlo simulations. Work in this direction has been carried out by the authors of Ref.~\cite{GENIE:VT}, who modified the GENIE event generator replacing the
nucleon energy and momentum distributions obtained from the RFGM with those predicted by realistic spectral 
functionss.  The resulting package, referred to as GENIE-$\nu$T, allows to take into account the strong energy-momentum correlation observed by 
$(e,e^\prime p)$ experiments at large missing energy and missing momentum, thus significantly improving the accuracy in the determination of neutrino energy. 

\acknowledgements

This research was supported by the Italian National Institute for Nuclear Research (INFN) under grant MANYBODY.

\end{document}